\documentclass[amsfonts, amssymb, amsmath, showkeys, nofootinbib, preprint,floatfix, prl, superscriptaddress]{revtex4-2}

\usepackage[colorinlistoftodos, color=green!40,prependcaption]{todonotes}
\usepackage{graphicx}% Include figure files
\usepackage{dcolumn}% Align table columns on decimal point
\usepackage{bm}% bold math
\usepackage{physics}
\usepackage{float}
\usepackage{xcolor}
\usepackage{amsmath}
\usepackage{amssymb}
\usepackage{units}
\usepackage[english]{babel}
\usepackage{placeins}
\usepackage{lipsum}
\begin{document}

\title{Spin wavepacket propagation in quasi-2D antiferromagnets}

\author{Yue Sun}
\affiliation {Department of Physics, University of California, Berkeley, California 94720, USA}
\affiliation {Materials Science Division, Lawrence Berkeley National Laboratory, Berkeley, California 94720, USA}

\author{Joseph Orenstein}
\affiliation {Department of Physics, University of California, Berkeley, California 94720, USA}
\affiliation {Materials Science Division, Lawrence Berkeley National Laboratory, Berkeley, California 94720, USA}

\begin{abstract}
Antiferromagnets are attractive platforms for the propagation of information via spin waves, offering advantages over ferromagnets in speed of response and immunity to external fields. A recent study of the quasi-2D antiferromagnet CrSBr reported that spin wavepackets propagate with group velocities that are orders of magnitude higher than expected from the magnon dispersion obtained by inelastic neutron scattering \cite{baeExcitoncoupledCoherentMagnons2022,scheieSpinWavesMagnetic2022}. Here we show that the anomalous magnitude and anisotropy of the group velocity, $v_g$, are naturally explained by considering the long-range magnetic dipole-dipole interaction. We also demonstrate that $v_g$ can be tuned over orders of magnitude by applying an external magnetic field or varying the sample thickness. Beyond spin wavepacket propagation, the dipolar interaction creates the possibility of non-equilibrium Bose-Einstein condensation in antiferromagnets, previously thought to be a property unique to ferromagnets. 
\end{abstract}

\maketitle

\section{Introduction}
Exploiting the electron spin degree of freedom is one of the central goals of condensed matter physics. An exciting direction is coupling spin to charge and lattice degrees of freedom to provide interconnections in hybrid quantum systems \cite{lachance-quirionEntanglementbasedSingleshotDetection2020}. To this end, it is essential to understand and control the generation, propagation, and detection of spin information. Recent progress in magnetically ordered systems has shown the promise of using spin waves -- collective excitations of the electron spins -- to transport information over large distances \cite{pirroAdvancesCoherentMagnonics2021,cornelissenLongdistanceTransportMagnon2015,lebrunTunableLongdistanceSpin2018,lebrunLongdistanceSpintransportMorin2020,hanBirefringencelikeSpinTransport2020,weiGiantMagnonSpin2022}. Antiferromagnets (AFMs) stand at the frontier of such research, given their promise of rapid response times and insensitivity to stray magnetic fields \cite{jungwirthAntiferromagneticSpintronics2016,baltzAntiferromagneticSpintronics2018}. 

Increasingly, attention has focused on quasi-two dimensional (2D) AFMs in which planar ferromagnetic order alternates in direction from layer to layer \cite{xingMagnonTransportQuasiTwoDimensional2019}. Of particular interest are AFMs in which magnetocrystalline anisotropy favors alignment of spin parallel to the layers \cite{lebrunTunableLongdistanceSpin2018,lebrunLongdistanceSpintransportMorin2020,hanBirefringencelikeSpinTransport2020,hoogeboomNonlocalSpinSeebeck2020}. Compared with easy-axis AFMs, such easy-plane AFMs exhibit highly tunable spin dynamics \cite{kalashnikovaImpulsiveGenerationCoherent2007,satohSpinOscillationsAntiferromagnetic2010,tzschaschelUltrafastOpticalExcitation2017} and potentially exhibit a form of dissipationless spin transport known as spin superfluidity \cite{soninSpinCurrentsSpin2010,soninSuperfluidSpinTransport2020}. 

A recent study of spin propagation performed on the easy-plane  antiferromagnet CrSBr \cite{baeExcitoncoupledCoherentMagnons2022} illustrates the potential of this class of materials. Bae et al. \cite{baeExcitoncoupledCoherentMagnons2022} demonstrated that the propagation of coherent spin waves, with typical energies below 1 meV, can be probed at visible wavelengths through their interaction with excitons -- a result with implications for the transduction of quantum information. However, the measured group velocity, $v_g$, of spin wavepackets presented a puzzle, as it was found to be orders of magnitude larger than the velocity estimated by extrapolating the magnon dispersion seen by neutron scattering \cite{scheieSpinWavesMagnetic2022} to the long wavelengths that are probed optically.  As a possible explanation, it was proposed that through magnetoelastic interactions, magnons in CrSBr can propagate through interaction with acoustic phonons.
\begin{figure}[t]
\centering
\includegraphics[width=1.0\columnwidth]{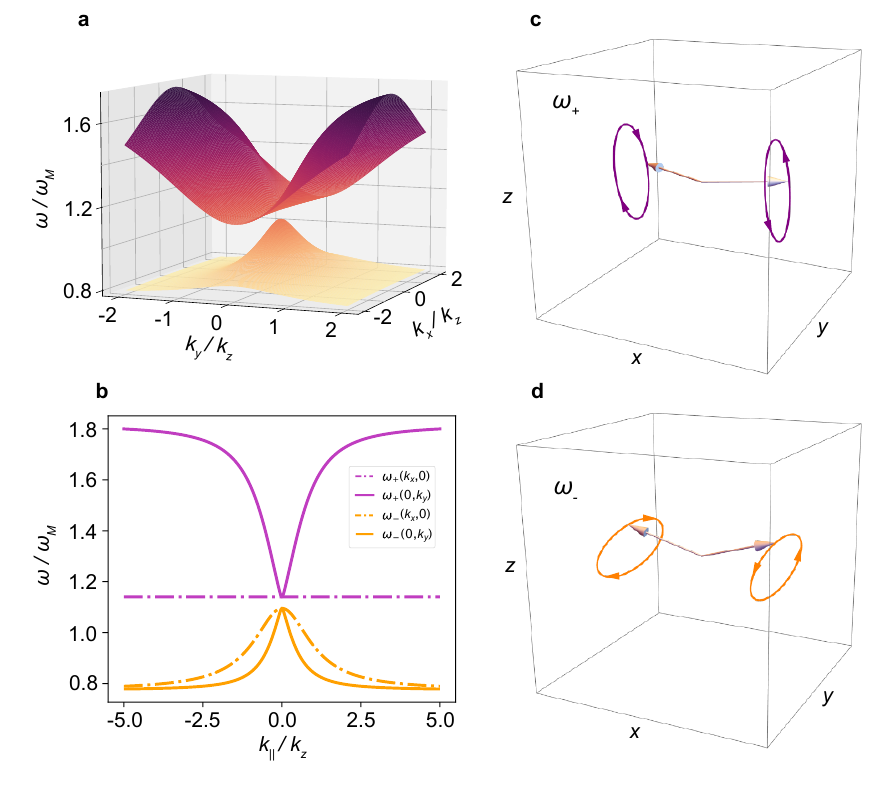}
\caption{(a) Spin wave frequencies for the two bands as a function of wavevector in the $k_x,k_y$ plane; (b) Energy dispersion along the $k_{x}$ axis (dash-dotted lines) and the $k_{y}$ axis (solid lines) for both $\omega_{-}$ (orange lines) and $\omega_{+}$ (purple lines) modes with $\omega_{T}/\omega_{M} = 1.0$, $\omega_{x}/\omega_{M} = 0.3$ and $\omega_{J}/\omega_{M} = 1.0$; (c,d) Normal mode eigenvectors of (c) $\omega_{+}$ and (d) $\omega_{-}$ modes for $k_{x}/k_{z} = 0$ and $k_{y}/k_{z}=3$.}
\label{fig:Fig1_dispersion}
\end{figure}

Here, we show that the puzzle is resolved without invoking coupling to phonons, and arises instead from the long-range magnetic dipole interaction. The dipolar coupling drastically alters the dispersion relations in the range of wavevectors probed optically, but not resolved by neutron scattering. Excellent agreement with the experimental results is obtained with a theory for dipole-coupled modes that follows directly from the semiclassical equations of motion for spin and Maxwell's equations. Our calculations of spin wave dispersion in the dipolar regime quantitatively account for all the the observed features of spin wavepacket propagation: the number of modes, the magnitude of their group velocities and the striking dependence of $v_g$ on propagation direction. Further, our theory has implications beyond resolving the puzzle presented by Ref. \cite{baeExcitoncoupledCoherentMagnons2022}.  We show that the dipolar dispersion depends strongly on sample thickness and magnetic field, enabling tuning $v_g$ over a large range. In addition, we show that including the exchange coupling between spins introduces a minimum in the dispersion relation, creating the possibility of a nonequilibrium Bose-Einstein condensate in an AFM.

\section{Wavepacket propagation in a biaxial antiferromagnet}
\subsection{Dispersion relation of dipolar modes}
Compared with ferromagnets, where dipolar effects have been well-recognized and studied \cite{damonMagnetostaticModesFerromagnet1961, hurbenTheoryMagnetostaticWaves1996, satohDirectionalControlSpinwave2012, demokritovBoseEinsteinCondensation2006,matsumotoTheoreticalPredictionRotating2011}, these interactions are often ignored in AFMs \cite{qaiumzadehSpinSuperfluidityBiaxial2017}, as there is no net magnetization in equilibrium. However, dipole interactions arise from dynamic fluctuations out of the equilibrium state, and are therefore relevant at long wavelengths in all forms of magnetic order. The collective modes in this regime are referred to as magnetostatic spin waves (MSWs) although they are fully dynamic; the term arises because their dispersion relations are obtained within the magnetostatic approximation, $\nabla\times \bm{H}=0$, which is valid because spin wave velocities are much smaller than the speed of light.

To solve for the normal modes at long wavelength, we model the easy plane AFM as alternating layers of magnetization, $\bm{M}_{1,2}$, which are antiparallel in equilibrium \cite{camleyLongWavelengthSurfaceSpin1980}. The Landau-Lifshitz equations for the dynamics of the magnetic sublattices,  
\begin{equation}
\label{LL}
    \frac{\partial \bm{M}_i}{\partial t}=- \gamma \bm{M}_i\times \bm{H}^{eff}_i,
\end{equation}

combined with Maxwell's equations in the magnetostatic regime, $\nabla \cdot \bm{B} =\nabla \times \bm{H} = 0$, form a closed set that yield the spin collective mode frequencies and eigenvectors. In Eq.~\ref{LL}, $\bm{H}^{eff}_i$ is the sum of effective fields arising from anisotropy and interlayer exchange, and the dynamical magnetic field $\bm{h}(t)$. The effective field on each sublattice is given by,
\begin{equation}
  \bm{H}^{eff}_i=-\frac{\partial F}{\partial \bm{M}_i}+\bm{h}_i(t), 
\end{equation}
where $F$ is the sum of exchange and anisotropy contributions, i.e. $F=F_{ex}+F_a$. The exchange term is,
\begin{equation}
\label{eq:exchange}
F_{ex}=J\frac{\bm{M}_{1}\cdot\bm{M}_{2}}{M_{s}^{2}},
\end{equation}
where $J$ is the exchange constant ($J>0$) and $M_{s}$ is the saturation magnetization of each sublattice. The easy-plane anisotropy is expressed as,
\begin{equation}
\label{eq:anisotropy}
F_{a} = K_{z}\frac{M_{1z}^{2}+M_{2z}^{2}}{M_{s}^{2}}-K_{x}\frac{M_{1x}^{2}+M_{2x}^2}{M_s^2},
\end{equation}
where $K_{x}$ and $K_{z}$ are anisotropy constants ($K_{x}, K_{z}>0$). The first term on the right-hand side of Eq. \ref{eq:anisotropy} confines the spins to the $xy$ plane while the second term expresses the easy-axis anisotropy within the plane. Although $K_x$ is typically much smaller than $K_z$, it plays an important role in the dispersion of long-wavelength spin waves.

Within an $x$-oriented domain the equilibrium magnetization is $\bm{M}_{1}=(M_{s},0,0)$ and $\bm{M}_{2}=(-M_{s},0,0)$ and small fluctuations from equilibrium are transverse, i.e. $\bm{m}_i=(m_{iy},m_{iz})$. Assuming solutions of the form $\bm{m}_i(\bm{r},t)=\bm{m_{i0}}e^{i(\bm{k}\cdot\bm{r}-\omega t)}$, we find two spin wave bands with dispersion given by,
\begin{widetext}
\begin{equation}
\label{eq:dispersion}
\omega_{\pm}(\bm{\hat{k}})=\sqrt{
    \omega_{T}\omega_{x}+\frac{\omega_{J}(\omega_{T}+\omega_{x})}{2}+\frac{\omega_{M}(k_{y}^{2}\omega_{T}+k_{z}^{2}\omega_{x})}{k^{2}}
    \pm A(\hat{\bm{k}})
    },
\end{equation}
\begin{equation}
\label{A(k)}
    A(\hat{\bm{k}})= \sqrt{\omega_{J}^{2}(\omega_{T}-\omega_{x})^{2}+4\omega_{J}\omega_{M}(\omega_{T}-\omega_{x})\left(\frac{\omega_{T}k_{y}^{2}-\omega_{x}k_{z}^{2}}{k^2}\right)+4\omega_{M}^{2}\left(\frac{\omega_{T}k_{y}^{2}+\omega_{x}k_{z}^{2}}{k^2}\right)^{2}},
\end{equation}
\end{widetext}
where $\omega_{M}=4\pi\gamma M_{s}$, $\omega_{x}=2\gamma K_x/M_s$, $\omega_{T}=2\gamma(K_{x}+K_{z})/{M_{s}}$,  $\hat{\bm{k}}=\bm{k}/k$ and $\gamma$ is the gyromagnetic ratio (see Supplemental Information for derivation of the dispersion relations). The spin wave frequencies depend only on the direction of the wavevector $\bm{k}$ and not its magnitude -- a property that arises from the long wavelength nature of the interaction \cite{camleyLongWavelengthSurfaceSpin1980}. 

To consider propagation in the plane of thin films or flakes, we focus on the dispersion of $\omega_{\pm}$ in the $k_x,k_y$ plane, as shown in Fig.~\ref{fig:Fig1_dispersion}a. Fig.~\ref{fig:Fig1_dispersion}b shows the dispersion along the $k_x=0$ and $k_y=0$ directions. It is clear from the dispersion relations that the propagation of spin waves in the two bands will have qualitatively different properties. The lower energy band disperses in all directions in the plane and is ``backwards moving'' in the sense that its phase and group velocity are opposite in sign. By contrast, the higher energy band is strongly dispersive and forward-moving for propagation in the $y$ direction (perpendicular to the equilibrium N\'{e}el vector), but exhibits no dispersion in the direction parallel to the equilibrium N\'{e}el vector. Finally, Figs.~\ref{fig:Fig1_dispersion}c and d illustrate the spin precession that characterize the $\omega_{-}$ and $\omega_{+}$ modes, respectively. In the regime where $K_z\gg K_x$, the $\omega_{+}$ mode (Fig.~\ref{fig:Fig1_dispersion}c) oscillates mainly within the $xz$ plane, while the $\omega_{-}$ mode (Fig.~\ref{fig:Fig1_dispersion}d) oscillates mainly within the $xy$ plane. 

%The similarity of normal modes between dipolar and exchange magnons indicates there can be a smooth crossover between them. We will discuss this crossover later.

\begin{figure}[t]
\centering
\includegraphics[width=0.9\columnwidth]{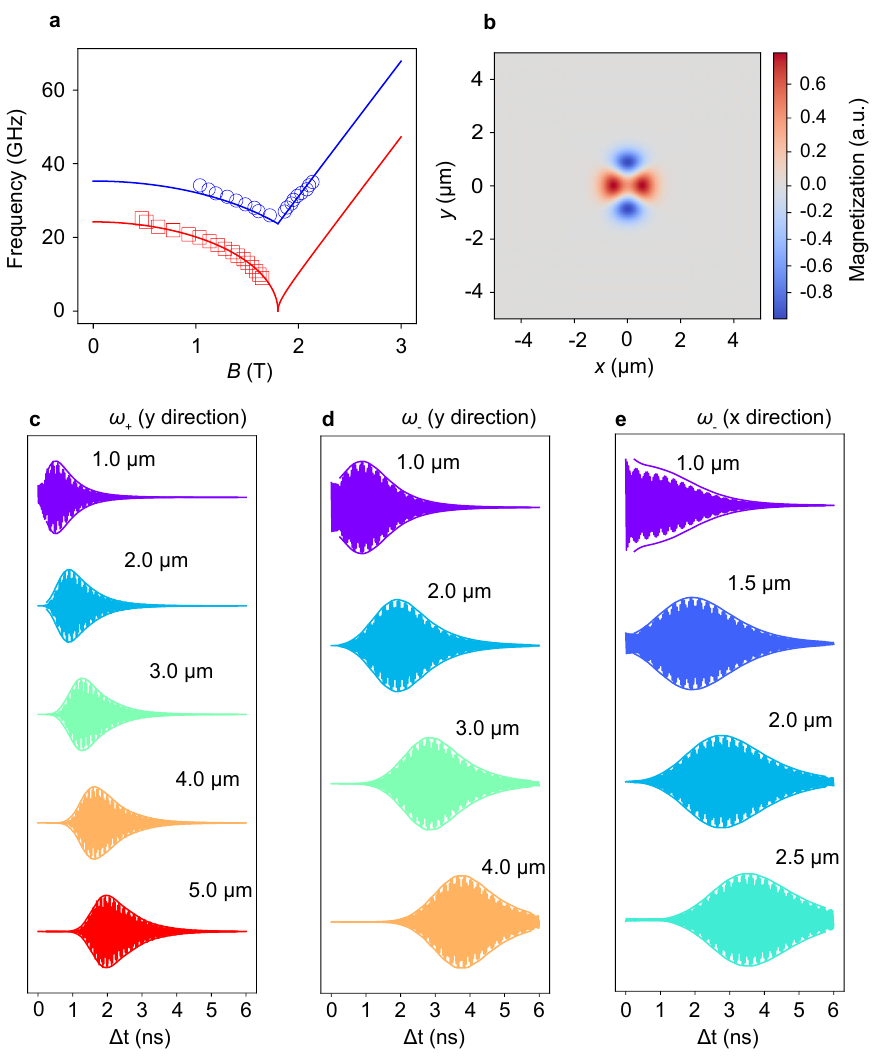}
\caption{(a) Spin wave frequencies as a function of out-of-plane magnetic field. The solid curves are fits to the dipolar magnon model. Data points are from Ref.~\cite{baeExcitoncoupledCoherentMagnons2022}; (b) Calculated spatial snapshot of spin wave propagation at a time delay of 109.8 ps; (c-e) Simulated magnon signal probed at several pump-probe separations for (c) the $\omega_{+}$ mode along the y direction, and (d,e) the $\omega_{-}$ mode along the y and x directions, respectively.}
\label{fig:Fig2_simulation}
\end{figure}

\subsection{Wavepacket propagation in CrSBr}
Below we show that the spin wave dispersion relations described above account quantitatively for wavepacket propagation in CrSBr. Photoexcitation launches a spin wavepacket comprised of plane waves whose wavevectors have a Gaussian distribution in the $k_x,k_y$ plane. We assume the thin film limit in which the pump penetration depth is much greater than the thickness, $d$, such that predominately modes with $k_z=\pi/d$ are excited. The wavepacket evolves in time according to, 
\begin{equation}
\label{eq:Integration}
\bm{m}_i(x,y,t)=\bm{m}_i e^{-\alpha t}\int d^{2}\bm{k}\ e^{-k^{2}\sigma^{2}/2} e^{-i\omega_{\pm}(\bm{k})t} e^{i\bm{k}\cdot \bm{r}},
\end{equation}
where $\bm{m}_i(x,y,t)$ is the dynamic local change in magnetization, $\alpha$ is the damping rate, $\bm{k}=(k_{x},k_{y})$, $\bm{r}=(x,y)$ and $\sigma$ is the laser spot size. Evaluating the integral in Eq. \ref{eq:Integration} requires the free energy parameters: saturation magnetization, anisotropy, and exchange terms; as well as the experimental parameters, $d$ and $\sigma$. The free energy parameters for CrSBr were determined from the magnetic field dependence of the spin wave frequency \cite{baeExcitoncoupledCoherentMagnons2022} which is replotted in Fig. \ref{fig:Fig2_simulation}a. The solid lines are fits obtained with $K_{x} = 1.9\times 10^{4}$ J/m$^{3}$, $K_{z} = 1.05\times 10^{5}$ J/m$^{3}$, $J = 5.6\times 10^{4}$ J/m$^{3}$ and $M_{s}=2.0\times 10^{5}$ A/m, values that are slightly different from those given in Ref.~\cite{baeExcitoncoupledCoherentMagnons2022} as our theory includes the demagnetization fields arising in the slab geometry. For the spot size we use the quoted value of 1.4 $\mu$m \cite{baeExcitoncoupledCoherentMagnons2022}. The only parameter entering the calculation not determined directly from the experiment was the sample thickness, $d$, for which a best fit value of 200 nm was obtained.   

Bae et al.~\cite{baeExcitoncoupledCoherentMagnons2022} presented features of the spin wave propagation in CrSBr following pulsed photexcitation in two formats. First, a map of magnetization in the $xy$ plane at a fixed time delay after photoexcitation revealed the anisotropic nature of the propagation. Second, resolving the wavepacket amplitude as a function of time for several distances from the pump focus enabled measurement of spin wave velocities along the principal axis directions. For comparison, Fig. \ref{fig:Fig2_simulation}b shows that magnetization at $t=100$ ps calculated using Eq. \ref{eq:Integration} captures the anisotropic character of the map reported in Ref. \cite{baeExcitoncoupledCoherentMagnons2022}. 

The spatial distribution of the magnetization shown in Fig. \ref{fig:Fig2_simulation}b clearly identifies two principal axes of propagation.  Figs.~\ref{fig:Fig2_simulation}c-e illustrate wavepacket propagation calculated for both directions for each of the spin wave branches, and may be compared directly with the data shown in Fig. 3 of Ref. \cite{baeExcitoncoupledCoherentMagnons2022}. The magnitude of the group velocities and their distinctive anisotropy are in excellent agreement with the results in CrSBr: the $\omega_{+}$ mode propagates only along the $y$ direction with velocity 3.0 km/s, while the $\omega_{-}$ mode propagates in both $y$ and $x$ directions, with velocities 1.0 km/s and 0.7 km/s, respectively. 

\subsection{Further predictions of the model}

\begin{figure}[t]
\centering
\includegraphics[width=1.0\columnwidth]{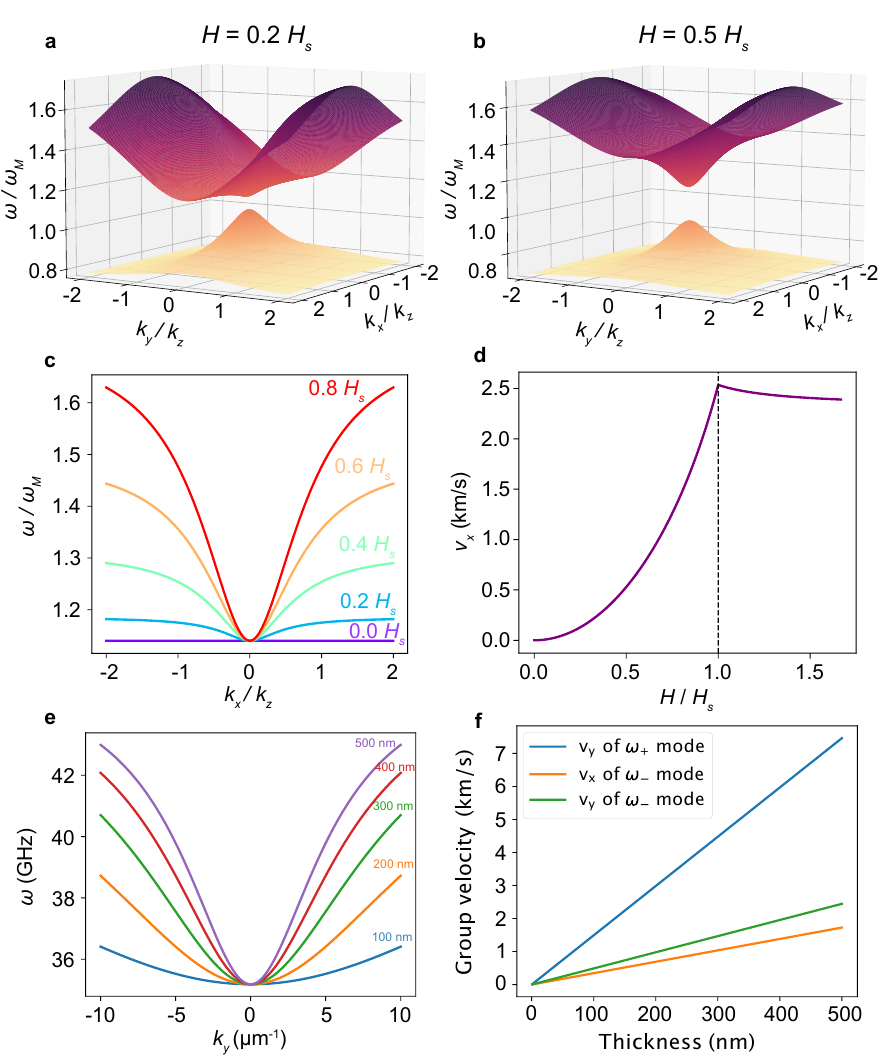}
\caption{(a,b) Spin wave dispersion for (a) out-of-plane external field $H=0.2 H_{s}$ and (b) $H=0.5 H_{s}$; (c) Linecuts of the $\omega_{+}$ mode dispersion along the $k_{x}$ direction at external fields ranging from zero to $0.8 H_{s}$; (d) Group velocity $v_{x}$ for the $\omega_{+}$ mode as a function of external field; (e) Linecuts of the $\omega_{+}$ mode dispersion along the $k_{y}$ direction for several sample thicknesses ranging from 100 nm to 500 nm; (f) Group velocities $v_{y}$ for the $\omega_{+}$ mode and $v_{x}$ and $v_{y}$ for the $\omega_{-}$ mode as a function of the sample thickness.}
\label{fig:Fig3_vx}
\end{figure}

Having shown that the dispersion of magnetostatic spin waves in a biaxial AFM resolves some puzzling observations in CrSBr \cite{baeExcitoncoupledCoherentMagnons2022}, we discuss several predictions that follow from our theory. We find that $v_g$ can be tuned over a broad range by a magnetic field applied perpendicular to the layers and by varying the film thickness. Figs.~\ref{fig:Fig3_vx}a and \ref{fig:Fig3_vx}b show the dispersion of the two spin wave bands in the $k_x,k_y$ plane for fields 0.2$H_s$ and 0.5$H_s$, respectively ($H_s$ is the saturation field). The dispersion of the $\omega_{+}$ mode in the $k_x$ direction, which was flat in zero field, is especially sensitive to applied fields, as is illustrated in Fig.~\ref{fig:Fig3_vx}c. Plotted in Fig.~\ref{fig:Fig3_vx}d are the maximum values of $v_g$ in the x direction as a function of field, which increase by orders of magnitude until reaching a maximum at $H_s$. 

Sensitivity of the dispersion relations to sample thickness arises when the penetration depth of pump photons is comparable or larger than $d$. In  this regime the spin wavepacket is comprised predominately of states in the $k_z=\pi/d$ subband. Fig. \ref{fig:Fig3_vx}e shows $\omega(k_x=0,k_y)$ for values of $k_z$ that correspond to thicknesses ranging from 100-500 nm. Because the dispersion is a function of $k_{x}/k_{z}$ and $k_{y}/k_{z}$, the group velocity in the $xy$ plane is proportional to $1/k_{z}$ and therefore to $d$. This dependence is illustrated in Fig.~\ref{fig:Fig3_vx}f, where we plot the largest velocity in a subband as a function of thickness.

Finally, we note that the negative group velocity of the $\omega_{-}$ mode creates the possibility of Bose-Einstein condensation (BEC) of magnons in an AFM. A nonequilibrium BEC has been observed in a ferrimagnet where the combination of magnetostatic and exchange interactions leads to a global minimum in the spin wave dispersion relation \cite{demokritovBoseEinsteinCondensation2006}. As the $\omega_{-}$ band in a biaxial AFM also shows negative group velocity at long wavelength, we anticipate that a global energy minimum will be found when exchange coupling between neighboring spins is included.  To test this expectation, we considered a cubic lattice of spins with nearest-neighbor interactions that are ferromagnetic within a layer and antiferromagnetic for spins in adjacent layers (See Supplemental Information for detailed derivation). Fig.~\ref{fig:Fig4_BEC}a shows $\omega_{-}(k_x,k_y)$ obtained in the presence of both dipolar and exchange coupling, revealing the global minimum in energy necessary to observe a magnon BEC. Fig.~\ref{fig:Fig4_BEC}b shows $\omega_{-}(k_x=0,k_y)$ and $\omega_{+}(k_x=0,k_y)$ in the first Brillouin zone, illustrating the crossover from dipolar to exchange dominated regimes.

\begin{figure}[t]
\centering
\includegraphics[width=1.0\columnwidth]{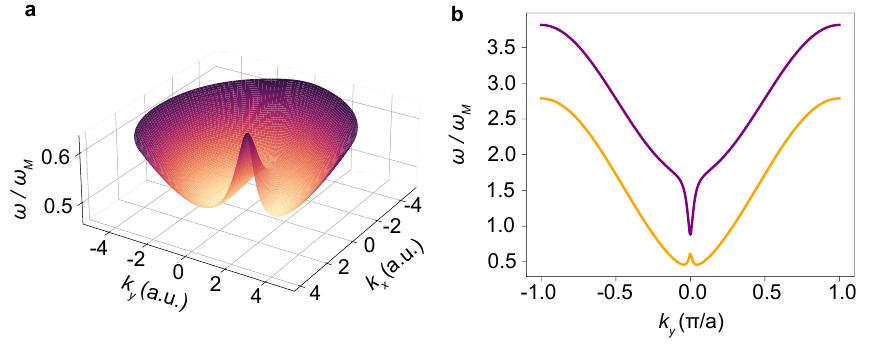}
\caption{(a) Energy dispersion of the $\omega_{-}$ mode including dipole-dipole and exchange interactions. The magnon mode has global energy minima at nonzero wavevector; (b) Energy dispersion of the $\omega_{-}$ and $\omega_{+}$ modes along the $k_{y}$ axis.}
\label{fig:Fig4_BEC}
\end{figure}

In conclusion, our work demonstrates that the propagation of optically generated spin wavepackets in a biaxial AFM is mediated by the long-range magnetic dipole interaction. Our model naturally explains the puzzling disparity between the group velocity observed in CrSBr and the value expected by extrapolating neutron scattering measurements to long wavelength. Also explained is striking dependence of $v_g$ on direction for each of the two spin wave bands. Finally, we described testable predictions of our theory: the spin wavepacket group velocity can be controlled by varying film thickness or applied magnetic field, and a nonequilibrium magnon BEC can be realized in biaxial AFMs.

\vspace{5 pt}
\begin{acknowledgments} 
We acknowledge support of the Quantum Materials program under the Director, Office of Science, Office of Basic Energy Sciences, Materials Sciences and Engineering Division, of the U.S. Department of Energy, Contract No. DE-AC02-05CH11231. J.O and Y.S received support from the Gordon and Betty Moore Foundation's EPiQS Initiative through Grant GBMF4537 to J.O. at UC Berkeley.
\end{acknowledgments}

\bibstyle{apsrev4-1}

\bibliography{Main_text_citation}

\end{document}

\title{Spin wavepacket propagation in quasi-2D antiferromagnets: supplementary information}
\maketitle

\section{Derivation of the dipolar magnon dispersion}
\label{zero_field}
In the orthorhombic crystal CrSBr, the magneto-crystalline anisotropy can be described as the combination of a hard axis along the out-of-plane direction (z-axis) and a easy axis along an in-plane direction (x-axis). The magnetization of each sublattice is denoted as $\bm{M}_{1}=\left(M_{1x}, M_{1y}, M_{1z}\right)$ and $\bm{M}_{2}=\left(M_{2x}, M_{2y}, M_{2z}\right)$. The free energy of the anisotropy, $F_{a}$, can be written as,
\begin{equation}
F_{a} = K_{z}\frac{M_{1z}^{2}+M_{2z}^{2}}{M_{s}^{2}}-K_{x}\frac{M_{1x}^{2}+M_{2x}^2}{M_s^2},
\end{equation}
where $K_{x}$ is the in-plane anisotropy energy, $K_{z}$ is the out-of-plane anisotropy energy ($K_{x}$, $K_{z}>0$) and $M_{s}$ is the saturation magnetization of each sublattice. The two sublattices are antiferromagnetically coupled, so the exchange energy is,
\begin{equation}
F_{ex}=J\frac{\bm{M}_{1}\cdot\bm{M}_{2}}{M_{s}^{2}},
\end{equation}
where $J$ is the exchange constant ($J>0$). The effective field $\bm{H}_{i}^{eff}$ is the sum of effective fields arising from anisotropy and interlayer exchange, and the dynamical magnetic field $\bm{h}(t)$. Therefore, our goal is to solve the Landau-Lifshitz equation for the magnon dispersion \cite{camleyLongWavelengthSurfaceSpin1980},
\begin{equation}
\label{eq:LL}
\begin{aligned}
&\frac{\partial \bm{M}_{i}}{\partial t}=-\gamma\bm{M}_{i}\times\bm{H}_{i}^{eff},
\\& \bm{H}_{i}^{eff} = -\frac{\partial (F_{a}+F_{ex})}{\partial \bm{M}_{i}} + \bm{h}(t),
\end{aligned}
\end{equation}
where $\gamma$ is the gyromagnetic ratio. 

Within an $x$-oriented domain the equilibrium magnetization is $\bm{M}_{1}=(M_{s},0,0)$ and $\bm{M}_{2}=(-M_{s},0,0)$ and small fluctuations from equilibrium are transverse, i.e. $\bm{m}_i=(m_{iy},m_{iz})$. We consider an infinitely large sample, so that the normal modes of magnetization and associated dynamics magnetic fields have the form of plane waves,
\begin{equation}
\label{eq:planewave}
\begin{aligned}
&m_{1y}(\bm{r},t) = m_{1y}e^{i(\bm{k}\cdot\bm{r}-\omega t)},
\ m_{1z}(\bm{r},t) = m_{1z}e^{i(\bm{k}\cdot\bm{r}-\omega t)},
\\&m_{2y}(\bm{r},t) = m_{2y}e^{i(\bm{k}\cdot\bm{r}-\omega t)},
\ m_{2z}(\bm{r},t) = m_{2z}e^{i(\bm{k}\cdot\bm{r}-\omega t)},
\\&h_{x}(\bm{r},t) = h_{x}e^{i(\bm{k}\cdot\bm{r}-\omega t)},
\\&h_{y}(\bm{r},t) = h_{y}e^{i(\bm{k}\cdot\bm{r}-\omega t)},
\\&h_{z}(\bm{r},t) = h_{z}e^{i(\bm{k}\cdot\bm{r}-\omega t)}.
\end{aligned}
\end{equation}
Substituting Eq.~\ref{eq:planewave} into Eq.~\ref{eq:LL}, we obtain the relation between $\bm{m}_{i}$ and $\bm{h}$,
\begin{equation}
\label{eq:m-h-relation}
\begin{aligned}
&\begin{bmatrix}
4\pi m_{1y} \\ 4\pi m_{1z}
\end{bmatrix}=
\begin{bmatrix}
\frac{\omega_{T}\omega_{M}}{(\omega_{x}+\omega_{J})\omega_{T}-\omega^{2}} & \frac{-i\omega\omega_{M}}{\omega_{x}(\omega_{T}+\omega_{J})-\omega^{2}} \\
\frac{i\omega\omega_{M}}{(\omega_{x}+\omega_{J})\omega_{T}-\omega^{2}} & \frac{\omega_{x}\omega_{M}}{\omega_{x}(\omega_{T}+\omega_{J})-\omega^{2}}
\end{bmatrix}
\begin{bmatrix}
h_{y} \\ h_{z}
\end{bmatrix},
\\& \begin{bmatrix}
4\pi m_{2y} \\ 4\pi m_{2z}
\end{bmatrix}=
\begin{bmatrix}
\frac{\omega_{T}\omega_{M}}{(\omega_{x}+\omega_{J})\omega_{T}-\omega^{2}} & \frac{i\omega\omega_{M}}{\omega_{x}(\omega_{T}+\omega_{J})-\omega^{2}} \\
\frac{-i\omega\omega_{M}}{(\omega_{x}+\omega_{J})\omega_{T}-\omega^{2}} & \frac{\omega_{x}\omega_{M}}{\omega_{x}(\omega_{T}+\omega_{J})-\omega^{2}}
\end{bmatrix}
\begin{bmatrix}
h_{y} \\ h_{z}
\end{bmatrix},
\end{aligned}
\end{equation}
where
\begin{equation}
\omega_{x} = \frac{2\gamma K_{x}}{M_{s}}; \ \omega_{T} = \frac{2\gamma K_{T}}{M_{s}};\ K_{T} = K_{x} + K_{z}; \ \omega_{J} = \frac{2\gamma J}{M_{s}}; \ \omega_{M} = 4\pi\gamma M_{s}.
\end{equation}
With the magnetostatic approximation, $\nabla\times \bm{h} = 0$, we can express the dynamic magnetic field as the gradient of a scalar potential $\psi$,
\begin{equation}
\begin{aligned}
&\psi = a e^{i(\bm{k}\cdot\bm{r}-\omega t)},
\\& \bm{h} = \nabla \psi = i\bm{k}a e^{i(\bm{k}\cdot\bm{r}-\omega t)}.
\end{aligned}
\end{equation}
In the next step we substitute the relation $\bm{b} = \bm{h}+4\pi(\bm{m}_{1}+\bm{m}_{2})$ into Eq.~\ref{eq:m-h-relation} to obtain,
\begin{equation}
\begin{aligned}
&b_{x} = h_{x},
\\&b_{y} = \left[1+\frac{2\omega_{T}\omega_{M}}{(\omega_{x}+\omega_{J})\omega_{T}-\omega^{2}}\right]h_{y},
\\&b_{z} = \left[1+\frac{2\omega_{x}\omega_{M}}{\omega_{x}(\omega_{T}+\omega_{J})-\omega^{2}}\right]h_{z}.
\end{aligned}
\end{equation}
Now with the Maxwell equation,
\begin{equation}
\label{eq:Maxwell}
\nabla\cdot \bm{b}=0,
\end{equation}
Eq.~\ref{eq:m-h-relation} takes the form,
\begin{equation}
\frac{\partial^{2} \psi}{\partial x^{2}}+\left[1+\frac{2\omega_{T}\omega_{M}}{(\omega_{x}+\omega_{J})\omega_{T}-\omega^{2}}\right]\frac{\partial^{2} \psi}{\partial y^{2}}+\left[1+\frac{2\omega_{x}\omega_{M}}{\omega_{x}(\omega_{T}+\omega_{J})-\omega^{2}}\right]\frac{\partial^{2} \psi}{\partial z^{2}}=0.
\end{equation}
Recalling that
\begin{equation}
\psi = a e^{i(\bm{k}\cdot\bm{r}-\omega t)},
\end{equation}
we have
\begin{equation}
\label{eq:Laplacian}
k_{x}^{2}+\left[1+\frac{2\omega_{T}\omega_{M}}{(\omega_{x}+\omega_{J})\omega_{T}-\omega^{2}}\right]k_{y}^{2}+\left[1+\frac{2\omega_{x}\omega_{M}}{\omega_{x}(\omega_{T}+\omega_{J})-\omega^{2}}\right]k_{z}^{2}=0.
\end{equation}
By solving Eq.~\ref{eq:Laplacian}, we obtain the dispersion relation that results from the dipole interaction in the magnetostatic regime,
\begin{equation}
\omega_{\pm}(\bm{\hat{k}})=\sqrt{
    \omega_{T}\omega_{x}+\frac{\omega_{J}(\omega_{T}+\omega_{x})}{2}+\frac{\omega_{M}(k_{y}^{2}\omega_{T}+k_{z}^{2}\omega_{x})}{k^{2}}
    \pm A(\hat{\bm{k}})
    },
\end{equation}
where $A(\hat{\bm{k}})$ is 
\begin{equation}
\label{A(k)}
    A(\hat{\bm{k}})= \sqrt{\omega_{J}^{2}(\omega_{T}-\omega_{x})^{2}+4\omega_{J}\omega_{M}(\omega_{T}-\omega_{x})\left(\frac{\omega_{T}k_{y}^{2}-\omega_{x}k_{z}^{2}}{k^2}\right)+4\omega_{M}^{2}\left(\frac{\omega_{T}k_{y}^{2}+\omega_{x}k_{z}^{2}}{k^2}\right)^{2}}.
\end{equation}

\section{The dipolar magnon dispersion in an external field}
\label{field_dependence}
We now consider the effect of an external field applied along the direction of the hard axis ($z$-axis). The free energy becomes,
\begin{equation}
F=K_{z}\frac{M_{1z}^{2}+M_{2z}^{2}}{M_{s}^{2}}-K_{x}\frac{M_{1x}^{2}+M_{2x}^2}{M_s^2}+J\frac{\bm{M}_{1}\cdot\bm{M}_{2}}{M_{s}^{2}}-H_{z}(M_{1z}+M_{2z}).
\end{equation}
Minimizing the free energy yields the equilibrium magnetization $\bm{M}_{1}=(M_{s}\cos\theta,0,M_{s}\sin\theta)$ and $\bm{M}_{2}=(-M_{s}\cos\theta,0,M_{s}\sin\theta)$, where,
\begin{equation}
\theta=
    \begin{dcases}
        \arcsin\left(\frac{H_{z}}{H_{s}}\right), & 0\leqslant H_{z} \leqslant H_{s}\\
        \frac{\pi}{2}, & H_{z}>H_{s}
    \end{dcases}
\end{equation}
$H_{s}$ is the saturation field given by $H_{s} = 2(J+K_{x}+K_{z})/M_{s}$. The normal modes are precession in the plane orthogonal to the equilibrium magnetization, which has the form $\bm{m}_{1}=(-m_{a}\sin\theta,m_{1y},m_{a}\cos\theta)$ and $\bm{m}_{2}=(-m_{b}\sin\theta,m_{2y},-m_{b}\cos\theta)$. We apply the same approach as in Sec. \ref{zero_field} to determine the dispersion relation.  As the calculations become quite lengthy we include them in the form of the attached Mathematica Notebook 1.

\section{Layered antiferromagnet model in a cubic lattice}
The calculations described in Sections~\ref{zero_field} and~\ref{field_dependence} appear to consider the exchange interaction explicitly, but the energy dispersions obtained do not shows the behavior expected for exchange coupled spins. This comes about because we considered each sublattice as a macrospin without internal lattice structure. In this section, we consider a layered antiferromagnet on a cubic lattice with nearest-neighbor interactions that are ferromagnetic within a layer and antiferromagnetic for spins in adjacent layers. Treating the system classically, we write each spin as $\bm{M}_{mnp}$, where $m,n$ are in-plane indices and $p$ is the layer index. The free energy including both exchange interaction and anisotropy energy is,
\begin{equation}
F = \frac{J_{z}}{M_{s}^{2}}\sum_{\langle p,p'\rangle} \bm{M}_{mnp}\cdot \bm{M}_{mnp'} - \frac{J_{\parallel}}{M_{s}^{2}}\sum_{\langle m,m'\rangle}\sum_{\langle n,n'\rangle} \bm{M}_{mnp}\cdot \bm{M}_{m'n'p} - \frac{K_{x}}{M_{s}^{2}}\sum_{m,n,p}(M_{mnp}^{x})^{2} + \frac{K_{z}}{M_{s}^{2}}\sum_{m,n,p}(M_{mnp}^{z})^{2},
\end{equation}
where $J_{z}$ and $J_{\parallel}$ are interlayer and intralayer exchange, respectively ($J_{z}>0$, $J_{\parallel}>0$) and $K_{x}$ and $K_{z}$ are in-plane easy-axis and out-of-plane hard-axis anisotropy constant, respectively ($K_{x}>0$, $K_{z}>0$). With the dynamic field $\bm{h}$, the effective field at the site ($m,n,p$) is,
\begin{equation}
\begin{aligned}
\bm{H}_{mnp} &= -\frac{\partial F}{\partial \bm{M}_{mnp}}+\bm{h}_{mnp}
\\& = -\frac{J_{z}}{M_{s}^{2}}(\bm{M}_{mn(p+1)}+\bm{M}_{mn(p-1)})+\frac{J_{\parallel}}{M_{s}^{2}}(\bm{M}_{(m+1)np}+\bm{M}_{(m-1)np}+\bm{M}_{m(n+1)p}+\bm{M}_{m(n-1)p})
\\& + \frac{2K_{x}}{M_{s}^{2}}M_{mnp}^{x}\hat{x} - \frac{2K_{z}}{M_{s}^{2}}M_{mnp}^{z}\hat{z} + \bm{h}_{mnp}.
\end{aligned}
\end{equation}
The equilibrium state is $\bm{M}_{mn(2p)}=(M_{s},0,0)$ and $\bm{M}_{mn(2p+1)}=(-M_{s},0,0)$. Again assuming the normal modes have the form of plane waves, we have
\begin{equation}
\begin{aligned}
&M_{mn(2p)}^{y} = u_{y}e^{i(k_{x}ma+k_{y}na+2k_{z}pa-\omega t)}
,\ M_{mn(2p)}^{z} = u_{z}e^{i(k_{x}ma+k_{y}na+2k_{z}pa-\omega t)}
\\&M_{mn(2p+1)}^{y} = v_{y}e^{i(k_{x}ma+k_{y}na+k_{z}(2p+1)a-\omega t)}
,\ M_{mn(2p+1)}^{z} = v_{z}e^{i(k_{x}ma+k_{y}na+k_{z}(2p+1)a-\omega t)}
\\& h_{mnp}^{x} = h_{x}e^{i(k_{x}ma+k_{y}na+k_{z}pa-\omega t)}
,\ h_{mnp}^{y} = h_{y}e^{i(k_{x}ma+k_{y}na+k_{z}pa-\omega t)}
,\ h_{mnp}^{z} = h_{z}e^{i(k_{x}ma+k_{y}na+k_{z}pa-\omega t)}
\end{aligned}
\end{equation}
Substituting into the Landau-Lifshitz equation yields,
\begin{equation}
\label{LL_cubic}
\begin{aligned}
&i\omega u_{y} = \gamma M_{s}\left[\frac{(2J_{z}+4J_{\parallel}+2K_{x}+2K_{z})u_{z}+2J_{z}\cos(k_{z}a)v_{z}-2J_{\parallel}[\cos(k_{x}a)+\cos(k_{y}a)]u_{z}}{M_{s}^{2}}-h_{z}\right]
\\& i\omega v_{y} = \gamma M_{s}\left[\frac{-(2J_{z}+4J_{\parallel}+2K_{x}+2K_{z})v_{z}-2J_{z}\cos(k_{z}a)u_{z}+2J_{\parallel}[\cos(k_{x}a)+\cos(k_{y}a)]v_{z}}{M_{s}^{2}}+h_{z}\right]
\\&i\omega u_{z} = \gamma M_{s}\left[\frac{-(2J_{z}+4J_{\parallel}+2K_{x})u_{y}-2J_{z}\cos(k_{z}a)v_{y}+2J_{\parallel}[\cos(k_{x}a)+\cos(k_{y}a)]u_{y}}{M_{s}^{2}}+h_{y}\right]
\\&i\omega v_{z} = \gamma M_{s}\left[\frac{(2J_{z}+4J_{\parallel}+2K_{x})v_{y}+2J_{z}\cos(k_{z}a)u_{y}-2J_{\parallel}[\cos(k_{x}a)+\cos(k_{y}a)]v_{y}}{M_{s}^{2}}-h_{y}\right]
\end{aligned}
\end{equation}
Combining Eq.~\ref{LL_cubic} with the Maxwell equation $\nabla\cdot\bm{b} = \nabla\cdot[\bm{h}+4\pi(\bm{u}+\bm{v})]=0$ yields the magnon dispersion. As the formulae are quite lengthy, they are shown in the Mathematica Notebook 2 attached.

\bibstyle{apsrev4-2}

\bibliography{SI_text_citation}